\documentclass[twocolumn,a4paper]{article} 

\usepackage{amsmath}
\usepackage{nolta2025}
\usepackage{txfonts}
\usepackage{color}
\usepackage{natbib}
\usepackage{graphicx}
\usepackage{subcaption} 
\usepackage{float}

\begin{document}

\title{Ellipsoidal Filtration for Topological Denoising of Recurrent Signals}

\author{
  Omer Bahadir Eryilmaz${}^\dag$, Cihan Katar{}$^\ddag$ $^\star$, and Max A. Little{}$^\dag$}

\address{
\dag School of Computer Science, University of Birmingham \\
\ddag Electrics and Electronics Department, Turkish-German University\\
$ \star $ Electronics and Communication Engineering Department, Istanbul Technical University\\
[5pt]
Email: \email{obe851@student.bham.ac.uk}, \email{cihan.katar@tau.edu.tr, katar20@itu.edu.tr},
\email{maxl@mit.edu}
}

\maketitle
\orcid{
Omer Bahadir Eryilmaz: 0009-0006-2229-6254,
Cihan Katar: 0000-0003-1247-9082,
Max Little: 0000-0002-1507-3822
}

\begin{abstract}
We introduce ellipsoidal filtration, a novel method for persistent homology, and demonstrate its effectiveness in denoising recurrent signals. Unlike standard Rips filtrations, which use isotropic neighbourhoods and ignore the signal’s direction of evolution, our approach constructs ellipsoids aligned with local gradients to capture trajectory flow. The death scale of the most persistent $H_1$ feature defines a data-driven neighbourhood for averaging. Experiments on synthetic signals show that our method achieves better noise reduction than both topological and moving average filters, especially for low-amplitude components.

\end{abstract}

\section{Introduction}

Recurrent signals often exhibit periodic-like behavior, but with frequencies that change over time, resulting in trajectories that do not exactly repeat yet retain structured recurrence. Such signals commonly arise in natural systems~\citep{hernandez-lemus_topological_2024,vieten_kinematics_2020}.

Classical low-pass filters remove noise by averaging nearby time samples to suppress high-frequency components. However, when the frequency content varies over time, as in recurrent signals, time-domain filtering may distort the signal.

A more effective approach relies on reconstructing the signal in state space, where recurrent trajectories densely fill a toroidal structure \citep{gakhar_sliding_2024}. This geometric perspective enables spatial averaging based on trajectory shape rather than time alone.

Robinson \cite{robinson_topological_2016} proposed a topological low-pass filter that performs neighbourhood averaging in state space using $k$-nearest neighbours ($k$-NN).

In this work, we improve this idea by introducing a principled method for neighbourhood selection. Rather than relying on $k$-NN, we propose a \textit{flow-aware ellipsoidal filtration} that constructs anisotropic neighbourhoods aligned with the local signal gradient, capturing the underlying geometry more effectively than simple spherical neighbourhoods.

\subsection{Continuous-Time Smooth Dynamical Systems}

A dynamical system is defined by how the state of a system—its complete set of variables—evolves over time according to a set of governing equations. The \emph{state space} of such a system is a geometric representation in which each point corresponds to a unique state of the system. The evolution of a continuous-time dynamical system defines a \emph{flow}, which traces continuous trajectories through state space. These trajectories may eventually converge to subsets known as \emph{attractors}, which characterise the long-term behaviour of the dynamical system \citep{strogatz_nonlinear_2018}. In the context of this study, we consider a state space that is corrupted by noise.

We next introduce topological data analysis (TDA), which offers a framework for quantifying the geometric and topological features of attractors.

\subsection{Persistent Homology in Time-Series Analysis}

TDA \citep{carlsson_topology_2009, chazal_introduction_2021} provides methods for quantifying geometric and topological features—such as connected components, loops, and voids—in high-dimensional data. The primary tool, persistent homology (PH), captures these features across multiple scales.



Attractors are sampled as point clouds, with points representing system states corrupted by observation noise. Each isolated point initially forms a connected component, captured by the zero-dimensional homology group \(H_0\). To reveal higher-dimensional topological features, such as loops (\(H_1\)) and voids (\(H_2\)), a filtration method progressively connects points at increasing spatial scales. A common approach is the \emph{Rips} filtration, defined as follows:

Let \( X \) be a metric space with metric \( d \), and let \( \alpha \geq 0 \). The Rips complex \( \mathrm{R}(X, \alpha) \) is the simplicial complex with vertex set \( X \), where a finite subset \( \{x_0, x_1, \ldots, x_k\} \subset X \) forms a \( k \)-simplex if and only if \(d(x_i, x_j) \leq \alpha \; \text{for all }\;0 \leq i, j \leq k.\)

The filtration constructs a nested sequence of complexes:
\[
\emptyset \subseteq R_{\alpha_1}(X) \subseteq R_{\alpha_2}(X) \subseteq \cdots \subseteq R_{\alpha_k}(X), \quad \alpha_1 < \cdots < \alpha_k
\]

\begin{figure}[t]
\centering
\includegraphics[width=1\linewidth]{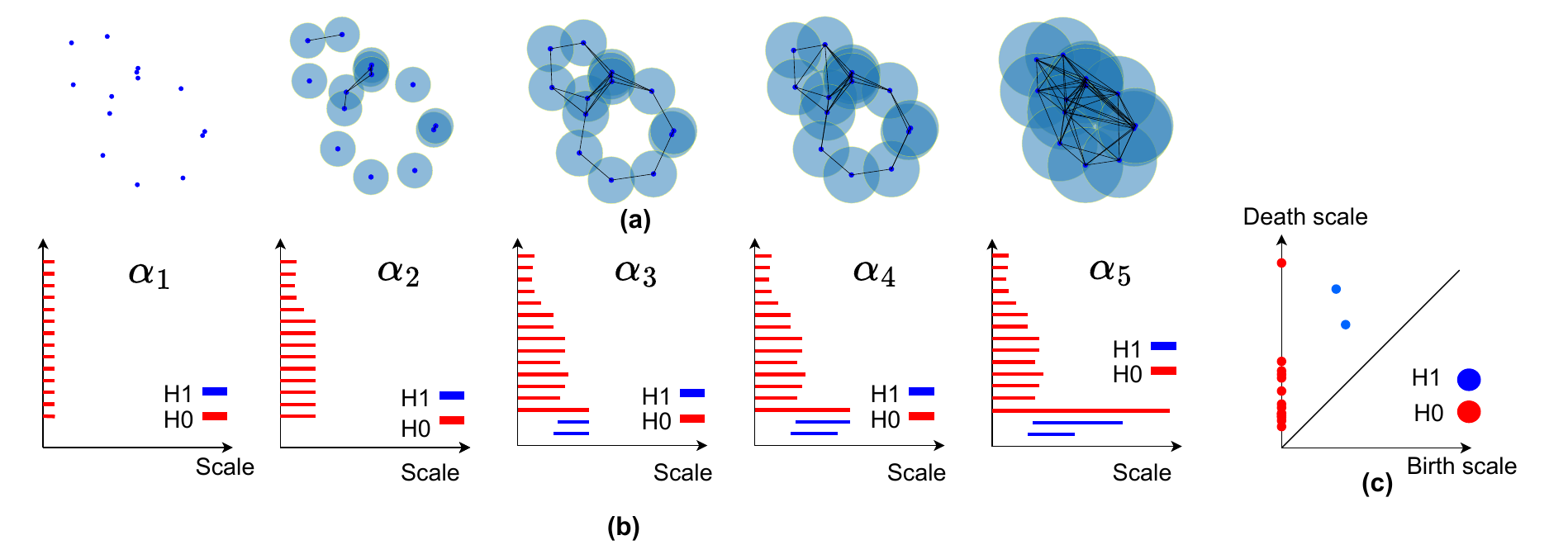}
\caption{Persistent homology: (a) Filtration process as the scale parameter \( \alpha \) increases; (b) Barcode representation of \( H_0 \) and \( H_1 \); (c) Persistence diagram (PD) for \( H_0 \) and \( H_1 \). By \( \alpha_2 \), some connected components have merged due to intersecting balls, forming edges. By \( \alpha_4 \), the small circular feature has already disappeared, as reflected in the barcode representations. The final barcode at \( \alpha_5 \) encodes the same topological information as the persistence diagram.}

\label{fig:persistent_homology}
\end{figure}

Persistent homology records when topological features appear (birth) and disappear (death) as the scale increases. The lifespan of each feature, defined as its persistence \( p = \alpha_d - \alpha_b \), is typically represented using persistence diagrams or barcodes (see Figure~\ref{fig:persistent_homology}).

\section{Method}
\label{sec:method}
\subsection{Ellipsoidal Filtration}

\subsubsection{Ellipsoid Construction and Alignment}

To analyse the local flow of the signal, we estimate gradients using a sliding-window averaging technique. For each point, the gradient is defined as the difference between the mean of the forward and backward windows, each of length \( N = 3 \). This symmetric formulation captures local directional trends while reducing the effect of noise.

At each point \( p \), we define an ellipsoid whose axes align with the local gradient direction.


\begin{figure}[!htbp]
    \centering
    \begin{subfigure}{0.48\linewidth}
        \centering
        \includegraphics[width=1\linewidth]{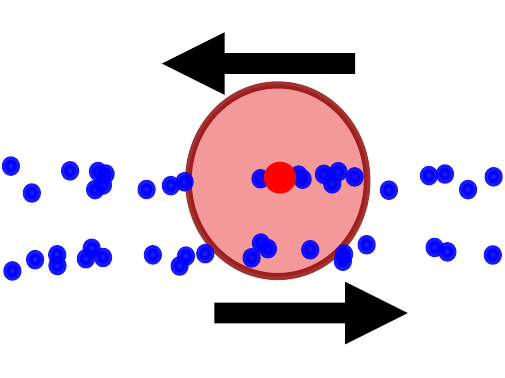}
        \caption{}
        \label{fig:filtation_circle}
    \end{subfigure}
    \hfill
    \begin{subfigure}{0.48\linewidth}
        \centering
        \includegraphics[width=1\linewidth]{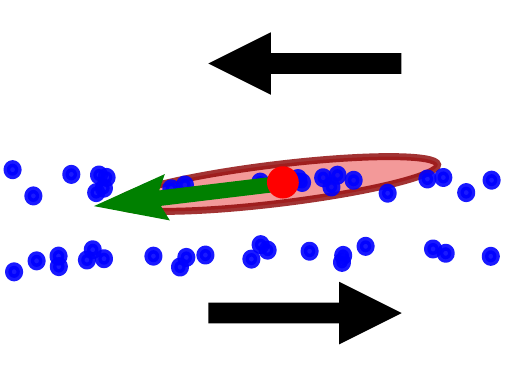}
        \caption{}  
        \label{fig:filtation_elips}
    \end{subfigure}
    
\caption{Filtration methods: (a) Spherical filtration; (b) Ellipsoidal filtration. The red dots denote the centres of the circle and ellipse. Black arrows indicate the global counter-clockwise data flow, while the green line shows the local flow direction at the centre of the ellipsoid. In (a), the circle includes points from opposing flow directions, which results in destructive averaging of small-scale signal components.}

    \label{fig:filtation_methods}
\end{figure}

Figure~\ref{fig:filtation_methods}, we compare two filtration methods. During spherical filtration, data with a specific direction can cause an intersection between two directions. However, the ellipsoidal filter adjusts the filter’s shape using data gradients to adapt to the data flow.
\subsubsection{Filtration via Ellipsoid Intersections}

Once the ellipsoids are constructed and aligned, we define a filtration by scaling them with a parameter \( \alpha \). As \( \alpha \) increases, ellipsoids expand, and intersections between them define edges in the simplicial complex. Higher-order simplices are formed by considering intersections among multiple ellipsoids.

To determine whether two ellipsoids \( E_1 = E(\Sigma_1, \mu_1) \) and \( E_2 = E(\Sigma_2, \mu_2) \) intersect, we employ the intersection condition proposed in \cite{gilitschenski_robust_2012}.

An ellipsoid \( E(\Sigma, \mu) \) in \( \mathbb{R}^n \) is defined as:
\begin{equation}
E(\Sigma, \mu) = \left\{ x \in \mathbb{R}^n \mid (x - \mu)^T \Sigma^{-1} (x - \mu) \leq 1 \right\}
\end{equation}
where:
\begin{itemize}
    \item \( \Sigma \in \mathbb{R}^{n \times n} \) is a symmetric positive definite shape matrix
    \item \( \mu \in \mathbb{R}^n \) is the center of the ellipsoid.
\end{itemize}

For two ellipsoids \( E_1 = E(\Sigma_1, \mu_1) \) and \( E_2 = E(\Sigma_2, \mu_2) \), their intersection is determined by the function:
\begin{equation}
K(s) = 1 - s (1 - s) (\mu_2 - \mu_1)^T \Sigma_2 E_s^{-1} \Sigma_1 (\mu_2 - \mu_1)
\end{equation}
where
\begin{equation}
E_s = s \Sigma_1 + (1 - s) \Sigma_2
\end{equation}

The convex function \( K(s) \) is defined over the interval \( s \in (0,1) \) and provides a necessary condition for the intersection of two ellipsoids. If \( K(s) < 0 \), the ellipsoids do not intersect. If there exists a unique \( s \) such that \( K(s) = 0 \), they touch at a single point. If \( K(s) > 0 \) for all \( s \), the ellipsoids overlap.

To efficiently test for intersection, we minimise \( K(s) \) over \( s \in (0,1) \) using golden-section search. If the minimum of \( K(s) \) is negative, the corresponding ellipsoids do not intersect and are discarded from further consideration.

We added each intersecting pair of vertices to the simplex tree data structure \citep{ boissonnat_efficient_2018} to construct the simplicial complex and compute persistence diagrams.

\subsection{Application of Ellipsoidal Filtration in Noise Removal}
\begin{figure}[!htbp]
    \centering
    \includegraphics[width=1\linewidth]{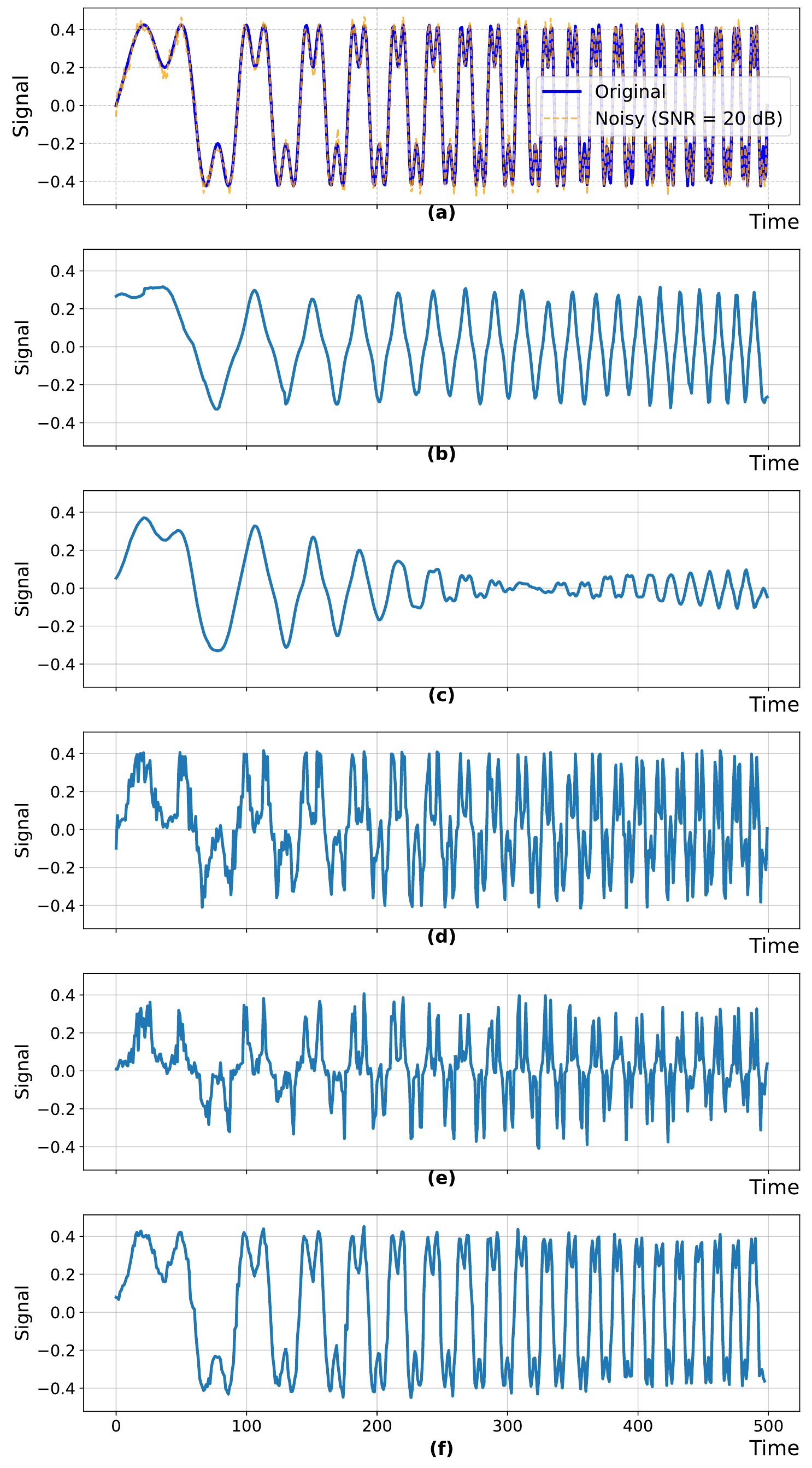}
    \caption{Comparison of filters applied to denoise the signal \( y(t) \) at SNR = 20~dB. (a) Original and noisy signals, (b) Adaptive moving average, (c) Moving average, (d) \(k\)-NN topological filter, (e) Spherical topological filter, (f) Ellipsoidal topological filter.}
    \label{fig:filtered_signals}
\end{figure}

For recurrent signals, our aim is to perform neighbourhood averaging in state space that reduces noise without distorting the signal. To guide this, we use the death scale of the most persistent \( H_1 \) feature, which reflects the size of the main loop in the attractor. This scale determines how large a neighbourhood to use for averaging.
Standard Rips filtration uses circular (isotropic) neighbourhoods that ignore the signal’s local direction. This can lead to averaging over unrelated parts of the trajectory, especially in low-amplitude regions, and may distort the signal—a limitation shown in Figure~\ref{fig:filtered_signals}.
In contrast, our ellipsoidal filtration defines neighbourhoods that follow the signal’s flow. This allows more relevant points to be included in each neighbourhood, improving noise reduction while preserving the signal’s structure.

\section{Experiments}
\subsection{Synthetic Data: Recurrent Signal}

To evaluate the performance of the proposed low-pass filters, we generated a 2D synthetic signal that simulates recurrent, nonstationary behavior. The signal evolves on a non-uniformly flattened toroidal trajectory with increasing frequency and anisotropic deformation. It is defined as:
\begin{equation}
x(t) = 10 \cos\big( \theta(t)t \big), \quad y(t) = 2 \sin\big( \theta(t)t \big) \cdot S(x(t))
\end{equation}
where the instantaneous angular frequency increases linearly over time:
\begin{equation}
\theta(t) = 2\pi \left( f_{\text{start}} + (f_{\text{end}} - f_{\text{start}}) \frac{t}{t_{\max}} \right),
\end{equation}
with \( f_{\text{start}} = 1 \) Hz, \( f_{\text{end}} = 10 \) Hz, and time \( t \in [0, 2] \) sampled uniformly with \( n = 500 \) points.

The squeeze function \( S(x) \) introduces anisotropy in the \( y \)-direction by attenuating amplitudes near the center of the trajectory:
\begin{equation}
S(x) = 1 - 0.9 \exp\left( -0.5 \left( \frac{x}{\max |x|} \right)^2 \right).
\end{equation}

To assess robustness, Gaussian noise was added at various signal-to-noise ratios (SNRs).

\subsection{Low-Pass Filters}
We applied five different low-pass filters: two classical filters, namely the moving average and the adaptive moving average filter, and three topological filters based on different neighborhood choices—\( k \)-NN, spherical, and ellipsoidal neighborhoods. 

For the ellipsoidal and spherical topological filters, the scale was set to the death time of the most persistent \( H_1 \) homology class, ensuring that the dominant cycle was captured while incorporating as many points as possible into the averaging process. The \( k \)-NN filter used \( k = 20 \) for a dataset of 500 points. 

For comparison, we also applied a moving average filter (also called a \emph{boxcar} filter when the weights are uniform) with a fixed window of 20 samples, and an adaptive version where the dominant frequency in each 100-sample segment was estimated using the Fast Fourier Transform, and the window size was adjusted according to the Nyquist criterion.

\section{Results}
We show that our topological low-pass filter, built on ellipsoidal filtration, is highly effective compared to other filters, particularly in recovering the small-amplitude components of the signal, as illustrated in Figure~\ref{fig:filtered_signals}.

\begin{figure}[!htbp]
    \centering

    \begin{subfigure}{\linewidth}
        \centering
        \includegraphics[width=1\linewidth]{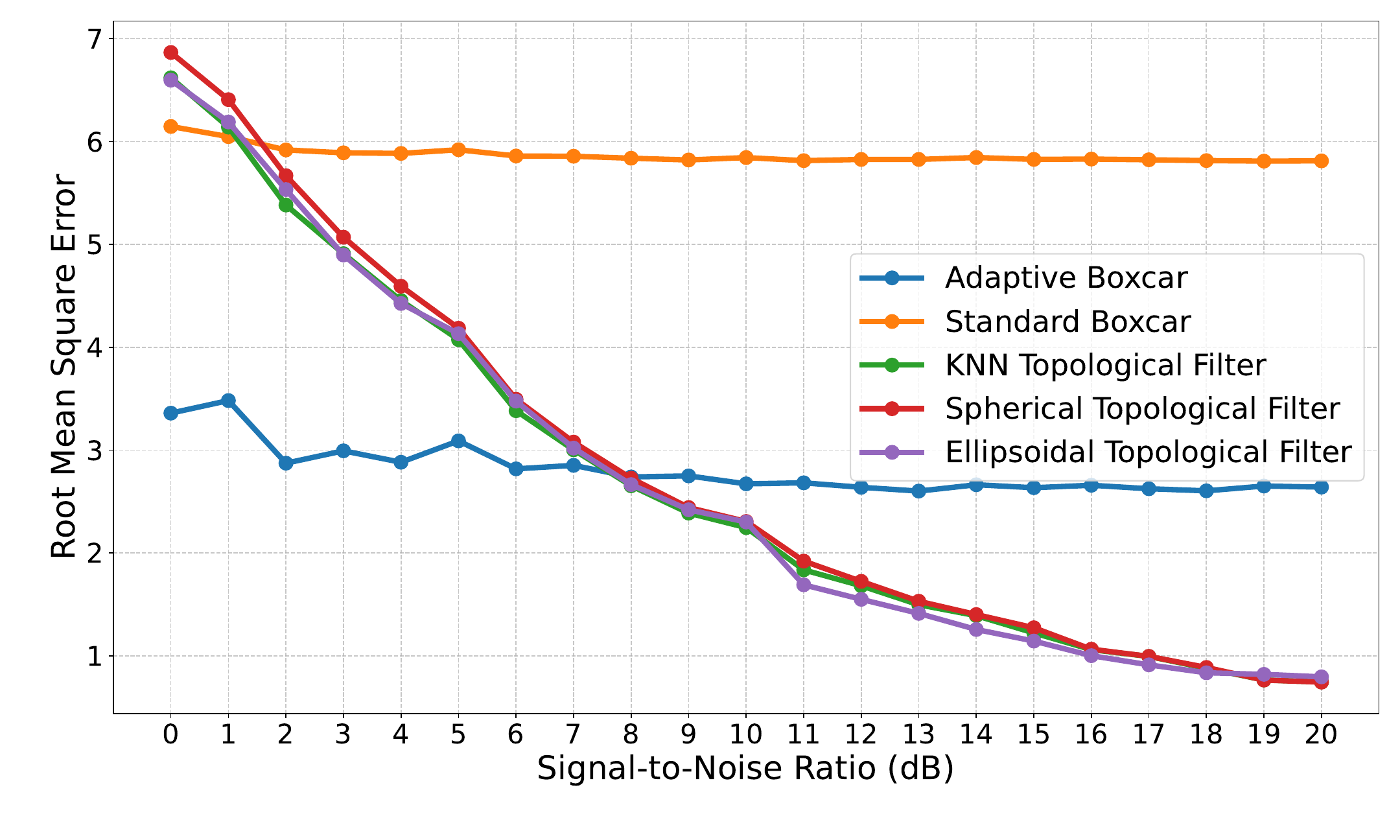}
        \caption{RMSE vs SNR for \(x(t)\) (high-amplitude component).}
        \label{fig:snr_rmse_x}
    \end{subfigure}

    \vspace{0.8em}

    \begin{subfigure}{\linewidth}
        \centering
        \includegraphics[width=1\linewidth]{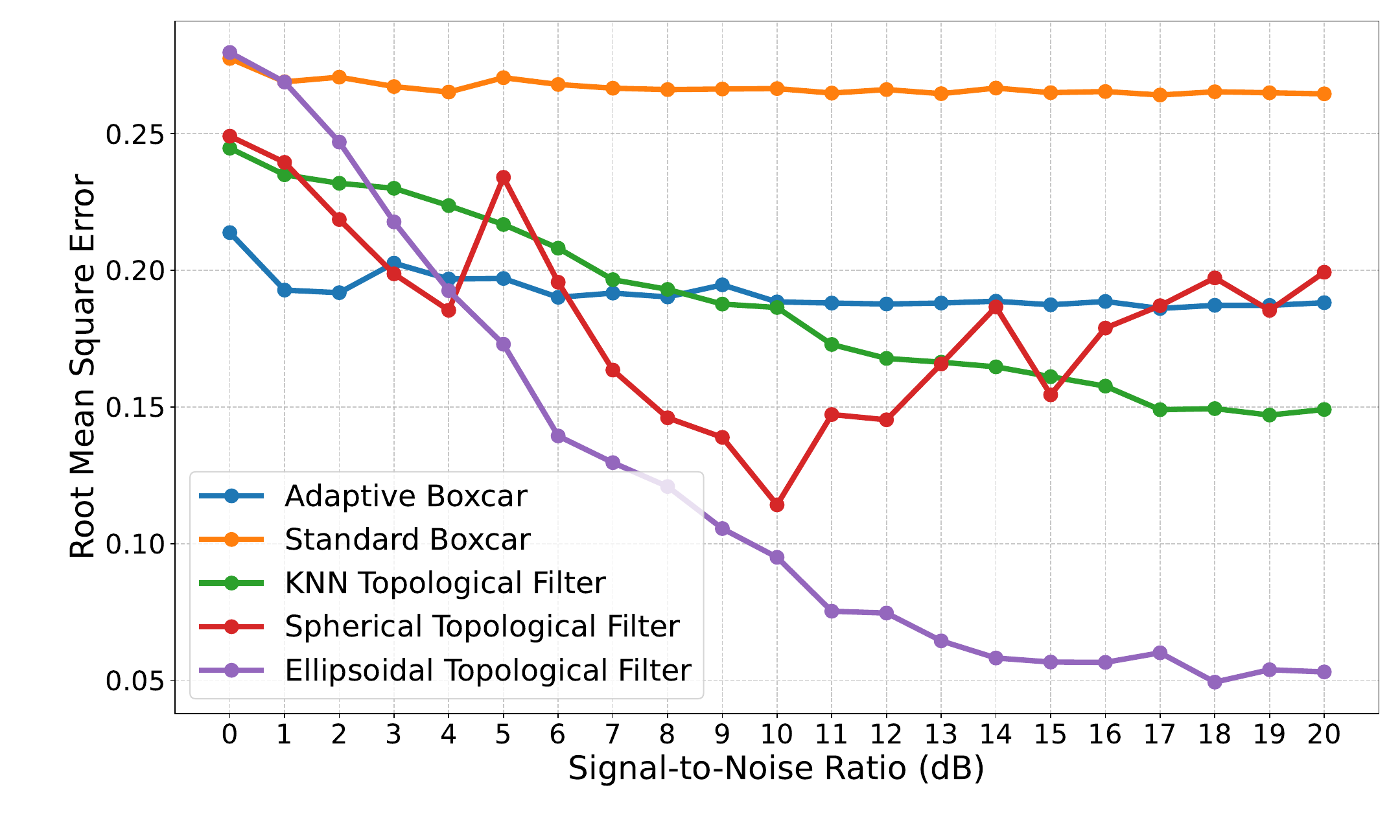}
        \caption{RMSE vs SNR for \(y(t)\) (low-amplitude component).}
        \label{fig:snr_rmse_y}
    \end{subfigure}

    \caption{Comparison of RMSE values across SNR levels for \(x(t)\) and \(y(t)\) using different filtration methods.}
    \label{fig:snr_rmse}
\end{figure}

As the SNR increases, topological filters outperform filters based on temporal windowing (see Figure~\ref{fig:snr_rmse}). This is because, although moving average filters smooth the signal, they struggle to preserve its structure—particularly when the signal's frequency varies. In contrast, topological filters perform spatial averaging in the state space (see Figure~\ref{fig:filtation_methods}), which is inherently more robust to frequency variation. However, this approach requires longer signals to ensure a sufficient number of close returns. These returns enable the inclusion of more structurally relevant samples in the averaging process, resulting in smoother and more accurate reconstructions.

In practice, close returns in the state space can occur before the completion of a full cycle \citep{vieten_kinematics_2020}. The ellipsoidal neighbourhood addresses this issue by avoiding destructive averaging of low-amplitude components through alignment with the local data flow, unlike uniform neighbourhoods that may include points from opposite directions (see Figure~\ref{fig:filtation_methods}). At the same time, the ellipsoidal approach incorporates enough relevant points to perform comparably to other neighbourhood choices when denoising the high-amplitude \( x(t) \) component, as shown in Figure~\ref{fig:snr_rmse_x}.

\section{Conclusion}
\label{sec:discussion}

We introduced an ellipsoidal filtration method for persistent homology and demonstrated that it improves neighbourhood selection by adapting to the local flow of the signal. This adaptation provides a data-driven neighbourhood for averaging, enabling more effective noise reduction while preserving the structure of the signal. In particular, it improves the recovery of low-amplitude components that are often suppressed by conventional filters.

Although topological filters achieve better performance in noise reduction for recurrent signals, their high computational cost may limit their applicability to high-dimensional time series. Future work should aim to improve the computational efficiency of these methods and evaluate their performance on real-world recurrent datasets. 

\section*{Acknowledgements}
Omer Eryilmaz gratefully acknowledges the financial support provided by the Ministry of National Education of Türkiye.
\bibliographystyle{unsrt}  
\bibliography{references} 

\end{document}